\documentstyle[12pt]{article}
\newcommand{\ald}{\dot \alpha}
\newcommand{\bed}{\dot \beta}
\newcommand{\gad}{\dot \gamma}
\newcommand{\ded}{\dot \delta}
\newcommand{\mud}{\dot \mu}

\newcommand{\dald}{{\bar{\partial}}_{\dot \alpha}}
\newcommand{\nald}{{\bar{\nabla}}_{\dot \alpha}}
\newcommand{\ned}{{\bar{\nabla}}_{\dot \beta}}

\newcommand{\ej}{\cal E}
\newcommand{\hej}{\hat {\ej}}
\newcommand{\pk}{ |p=\kappa^{J}}
\newcommand{\gpk}{g_{J-1 {\pk}}}
\newcommand{\gpi}{g^{-1}_{J-1 {\pk}}}
\begin{document}
\textwidth 160mm
\textheight 240mm
\topmargin -20mm
\oddsidemargin 0pt
\evensidemargin 0pt
\newcommand{\beq}{\begin{equation}}
\newcommand{\eeq}{\end{equation}}
\begin{titlepage}
\begin{center}

\huge{\bf SD Perturbiner in Yang-Mills+Gravity}

\vspace{1.5cm}

\large{\bf K.G.Selivanov }

\vspace{1.0cm}

{ITEP, B.Cheremushkinskaya 25, Moscow, 117259, Russia}

\vspace{1.9cm}

{ITEP-TH-59/97}

\vspace{1.0cm}

\end{center}
              

\begin{abstract}
An explicit self-dual classical solution of the type of
perturbiner in Yang-Mills theory interacting with gravity is obtained. 
This allows one to describe the tree self-dual gluonic form-factors  including any 
number of positive-helicity gravitons in addition to the positive
helicity gluons.
\end{abstract}
\end{titlepage}

\newpage
\setcounter{equation}{0}

\section{Introduction}
In this letter we continue our study of {\it perturbiners}
\cite{RS1},\cite{RS2}, \cite{RS3},  i.e.,
solutions of field equations (FE) which are generating functions for tree
amplitudes in the theory (more precisely - for tree form-factors,
that is ``amplitudes'' with a number of on-shell particles and one
off-shell particle in the coordinate representation). Such solution 
can be given an intrinsic definition which is formally independent
on the Feynman diagrams and which is universal in the sense that
it is applicable in any field theory. In words, the definition is as follows.
Take a linear combination of plane wave solutions of linearized
FE (with appropriate polarization factors etc.) so that
every plane wave is multiplied with a corresponding formal nilpotent
variable (a rudiment of the symbol of annihilation-creation operator).
The perturbiner is a (complex) solution of the (nonlinear) FE which is
polynomial in the given set of (nilpotent) plane waves, ${\ej}$, and whose first
order part is the given solution of the linearized FE. This definition
is actually nothing but rephrasing the usual Feynman perturbation
theory for the tree form-factors{\footnote{the nilpotency
is equivalent to considering only amplitudes with no particle having
identical quantum numbers}}. Nevertheless, this definition  
 appeared to be very convenient in the cases where FE can be treated
by some other powerful methods different from the perturbation theory.
This is actually the case of the self-duality (SD) equations, both
in  gravity and in Yang-Mills (YM) theory , allowing the use of the twistor
constructions \cite{Penrose}, \cite{Ward}. The way of reducing from the
generic perturbiner to the SD one is obvious: one includes into the
plane wave solutions of the linearized FE only SD plane waves, which
is equivalent to describing only amplitudes with on-shell particles of 
a given helicity (say, of the positive one). Miraculously, the twistor
construction works extremely efficiently in the case of perturbiners,
in a sense, more efficiently than in the case of instantons. In the instanton
case it gives only explicit description of the moduli
space of solutions \cite{ADHM}, while in the perturbiner case it gives
explicit description of fields (\cite{RS2}, \cite {RS3} and below).

In the case of YM theory, the idea to use the SD equations to describe
the so-called like-helicity amplitudes (see, e.g., the reviews \cite{mapa},
\cite{dixon})
was first formulated in \cite{Ba} and, independently, in \cite{Se}.
 In \cite{Ba} it was basically shown that  the SD
equations reproduce the recursion relations for tree like-helicity gluonic 
form-factors (also called ``currents''), obtained originally 
in ref.\cite{BG} from the Feynman diagrams; the  corresponding solution of 
SD equations was than obtained in terms of the known solutions 
refs.\cite{BG} of the recursion relations 
for the ``currents''. In ref.\cite{Se} an example of SD perturbiner was
obtained in the SU(2) case by a 'tHooft anzatz upon further restriction on the
on-shell particles included. A bit later an analogous solution was studied
in ref.\cite{KO}, where consideration was based
on solving  recursion relations analogous to refs. \cite{BG}.  
In \cite{RS2} the YM SD perturbiner was constructed by the twistor methods
which also allowed us to find its antiSD deformation, that is to add one
opposite-helicity gluon and thus to obtain a generating
function for the so-called maximally helicity violating Parke-Taylor 
amplitudes \cite{PT},
\cite{BG}. Also, in \cite{RS2} the SD perturbiner was constructed in
the background of an arbitrary instanton solution.
In \cite{RS3} we presented the gravitational SD perturbiner. Again, as in the YM
case, the twistor methods \cite{Penrose}, \cite{AHS}  gave absolutely explicit
description of the fields. Thus, we presented a generating function for
tree gravitational form-factors with all on-shell gravitons having the same
(positive) helicity. 
    
In this letter I present the perturbiner type SD solution in the YM theory
interacting with gravity.
Actually, after construction of the  SD perturbiner
in YM  \cite{RS2} and in gravity \cite{RS3}, the present case is almost
automatic. One first notice that the YM energy-momentum tensor vanishes
on the self-dual fields. This means that the gravity sector is  the same as
in the  pure gravity case \cite{RS3}. What concerns to the YM sector, its
gravitational dressing happens to be very simple. The YM fields are
the same as in \cite{RS2} functions but its variables, 
the plane waves ${\ej}$,  get
gravitationally dressed in a simple way, Eqs.(\ref{dressing}),(\ref{dressing1}).

In the section 2 I describe self-dual solutions of the linearized FE which
serve as a base point in the definition of perturbiner, then in section 3
I reproduce some necessary detailes of the solution
in the case of pure gravity \cite{RS3}. In section 4 I describe the
gravitational dressing of the gluons. Gravitational dressing of the 
Parke-Taylor amplitudes will be described elsewhere.

\section{The plane wave solution of the free FE}
It will be convenient to use the spinor notations. To construct the perturbiner, 
one first defines the plane wave solutions
of the linearized FE. The appropriate SD solution of the linearized gravity
reads ($e^{\alpha \ald}_{N}$ is a vierbein, the capital Latin letters like 
$N, M, L$ will stand for sets of gravitons) reads
\beq
\label{lintet}
e^{\alpha \ald}_{N}=e^{\alpha \ald}_{\emptyset}+\sum_{n{\in}N}
{\ej}_{n}\frac{q^{\alpha}_{n}\lambda^{\ald}_{n}}{(q^{n},\ae^{n})^{2}}
q_{\mu}^{n}\lambda_{\mud}^{n}dx^{\mu \mud}
\eeq
In the equation (\ref{lintet}) $e^{\alpha \ald}_{\emptyset}$ stands for the flat 
vierbein, 
$e^{\alpha \ald}_{\emptyset}=dx^{\alpha \ald}$, ${\ej}_{n}$ is the plane
wave,  ${\ej}_{n}=a_{n}e^{k^{n}_{\alpha \ald} x^{\alpha \ald}}$, $a_{n}$ is
the nilpotent symbol, $a_{n}^{2}=0$, $k^{n}_{\alpha \ald}$ is an (asymptotic)
four-momentum of the $n$-th graviton. $k^{n}$ is light-like and hence
it decomposes into a product of two spinors, 
$k^{n}_{\alpha \ald}=\ae^{n}_{\alpha}\lambda^{n}_{\ald}$.{\footnote{the reality 
of the four momentum in Minkowski space assumes that $\lambda_{\ald}=
{\bar {\ae}}_{\alpha}$}} 
As a consequence of the SD condition, the polarization tensor in 
Eq.(\ref{lintet}) contains the same dotted spinor $\lambda^{n}$ as in the 
four-momentum $k^{n}$. The other
spinor entering the polarization factor, $q^{n}_{\alpha}$, is a reference
spinor needed to define the polarization tensor, the on-shell gauge freedom
being $q^{n}_{\alpha}{\rightarrow}q^{n}_{\alpha}+{\ae}^{n}_{\alpha}$.
The factor in the
denominator is introduced for normalization, the brackets of the type of 
$(p,q)$
here and below mean contraction of two spinors with the 
$\epsilon$-tensor, $(p,q)=\epsilon^{\alpha \beta}p_{\alpha}q_{\beta}$.
$n$ numbers gravitons in the set $N$.

{A solution of the free (i.e. linearized) SD equation in the YM sector 
consisting of $N$ plane waves looks as follows}
\beq
\label{free1}
{A^{(1) J}_{\alpha \ald}=
\sum_{j}^{J} \epsilon^{+j}_{\alpha \ald}{\hej}_{j}}
\eeq
{where the sum runs over gluons, $J$ is the number of gluons,\\
$\epsilon^{+j}_{\alpha \ald}$ is a four-vector defining a polarization of the $j$-th gluon,
${\hej}^{j}=t_{j}{\ej}^{j}=t_{j}a_{j}e^{ik^{j}x}$,
$t_{j}$ is a matrix defining color orientation of the $j$-th gluon.
$k^{j}_{\alpha \ald}$, as a light-like four-vector, decomposes into a product of two
spinors}
${k^{j}_{\alpha \ald}=\ae^{j}_{\alpha} \lambda^{j}_{\ald}}$. 
{The polarization $\epsilon^{+j}_{\alpha \ald}$,
as a consequence of the linearized SD equations, also decomposes into a product
of spinors, such that the dotted spinor is the same as in the decomposition of momentum k,
$\epsilon^{+j}_{\alpha \ald}=
\frac{q^{j}_{\alpha} \lambda^{j}_{\ald}}{( \ae^{j},q^{j})}$
where normalization factor  is defined with use of a convolution 
$(\ae^{j},q^{j} )=\varepsilon^{\gamma \delta}\ae^{j}_{\gamma}q^{j}_{\delta}=
{\ae^{j}}^{\delta}{q^{j}}_{\delta}$. Indexes are raised and lowered with the
$\varepsilon$-tensors.}
{The free anti-SD equation would give rise to a polarization
$\epsilon^{-}_{\alpha \ald}=\frac{\ae_{\alpha} \bar{q}_{\ald}}
{( \lambda,\bar{q})}$
The auxiliary spinors $q_{\alpha}$ and $\bar{q}_{\ald}$ form together a four-vector
$q_{\alpha \ald}=q_{\alpha}\bar{q}_{\ald}$ usually called a reference momentum.
The normalization  was chosen so that}
${\epsilon^{+} \cdot \epsilon^{-}=\varepsilon^{\alpha \beta}\varepsilon^{\ald \bed}
\epsilon^{+}_{\alpha \ald} \epsilon^{-}_{\beta \bed}=-1}$

As it is defined in the introduction (see also \cite{RS1},\cite{RS2}, \cite{RS3}),
the perturbiner is a solution of (nonlinear) FE which is polynomial in
${\ej}_{n}, n{\in}N, {\ej}_{j}, j=1,J$  and whose linear in ${\ej}$'s part is as in 
Eqs.(\ref{lintet}),(\ref{free1}). 

\section{Gravitational SD perturbiner}
In the given orthonormal
basis $e^{\alpha \ald}=e^{\alpha \ald}_{\mu \mud}dx^{\mu \mud}, 
\alpha,\mu=1,2, \ald,\mud={\dot 1},{\dot 2}$ the curvature form,
${\bf R}^{\alpha \ald}_{\beta \bed}=
R^{\alpha \ald}_{\beta \bed \gamma \gad \delta \ded}e^{\gamma \gad}
e^{\delta \ded}$ decomposes into its SD and antiSD part as follows
\beq
{\bf R}_{\alpha \ald \beta \bed}={\bf R}^{+}_{\ald \bed}\epsilon_{\alpha \beta}
+{\bf R}^{-}_{\alpha \beta}\epsilon_{\ald \bed}
\eeq
where $\epsilon_{\alpha \beta}$ and $\epsilon_{\ald \bed}$ are the standard
antisymmetric tensors, the forms ${\bf R}^{+}_{\ald \bed}$ and 
${\bf R}^{-}_{\alpha \beta}$ are symmetric, and, as in most places below,
the indices are raised and lowered with the $\epsilon$-tensors.
The forms ${\bf R}^{+}$ and ${\bf R}^{-}$ are identified as SD
and antiSD components of the curvature form {\footnote{SD and antiSD 
components could also be defined with respect to another half of indices
of the curvature form, which is more or less equivalent to what is done 
above.}} Thus, the SD equations in the case of gravity read
\beq
\label{SD}
{\bf R}^{-}=0
\eeq
SD curvature form automatically obeys gravity equations
without matter (or with self-dual matter, as explained in the Introduction).

The self-duality condition Eq.(\ref{SD}) assumes that the monodromy group
acts trivially on the undotted spinors, hence there is a covariantly constant
spinor $p_{\alpha}$. Moreover, the spin-connection can be put in such a form 
that the spinor $p_{\alpha}$ is simply a constant spinor (antiSD part of the
spin-connection can be gauge away). Such a choice
of the spin-connection and, correspondingly, of the vierbein 
$e^{\alpha \ald}_{N}$ will be always assumed. Then forms 
$e^{\ald}_{N}=p_{\alpha}e^{\alpha \ald}_{N}$ are shown to define a 
covariantly constant
complex structure, and the two-form (actually, $(2,0)$-form)
\beq
\label{Om} 
\Omega_{N}=\epsilon_{\ald \bed}e^{\ald}_{N}{\wedge}e^{\bed}_{N} 
\eeq
is shown to be covariantly constant. The other way around, one can show that
every covariantly constant $(2,0)$-form defines a solution of the
SD equation.
 
Next we are looking for coordinates $z^{\ald}_{N}$ (and 
${\bar{z}}^{\ald}_{N}$) which are singular and of degree $1$
in the auxilliary ${\bf CP^{1}}$ space (parametrized by the spinor
$p^{\alpha}$) and which are the Darbough
coordinates for the $(2,0)$ form $\Omega_{N}$ which is regular and
of degree $2$ in the auxilliary ${\bf CP^{1}}$ space.
\beq
\Omega_{N}=\epsilon_{\ald \bed}dz^{\ald}_{N}{\wedge}dz^{\bed}_{N}
\eeq
Thus any meromorphic coordinates  $z^{\ald}_{N}$ subject to
condition of regularity of  the $(2,0)$ form $\Omega_{N}$ define
a solution of the SD equation (being regular and of degree $2$
the form $\Omega$ is necessary just bilinear, as it must be
according to its definition Eq.(\ref{Om}).

In the case of perturbiner
the coordinates $z_{N}$ are supposed to have an expansion of the form
\beq
\label{poly1}
z_{N}=z_{\emptyset}+u_{N} 
\eeq
where $u_{N}$ is polynomial in $\ej$ and the term of degree $1$ in the 
polynomial is defined by the solution in the linearized approximation 
Eq.(\ref{lintet}). 
\beq
\label{poly2}
u^{(1)}_{N}=\sum_{n{\in}N}\frac{(p,q_{n})^{2}{\lambda}_{n}}
{(p,{\ae}_{n})(q_{n},{\ae}_{n})^{2}}{\ej}_{n}
\eeq
(To obtain the above expression one first constructs the form $\Omega$
out of the tetrade Eq.(\ref{lintet}) and than decomposes it into the
Darbough coordinates, which is straightforward and easy in the first
in $\ej$ order.)

The regularity of $\Omega$ requires that
\beq
\label{residue1}
\epsilon_{\ald \bed}dz^{\ald}_{N{\setminus}n}|_{p={\ae}_{n}}
{\wedge}res_{n}dz^{\bed}_{N}=0
\eeq
($N{\setminus}n$ notation stands for the set $N$ without $n$th particle
and $res_{n}=res|_{p={\ae}_{n}}$ means the residue at point $p={\ae}_{n}$.)
The condition (\ref{residue1}) implies that 
\beq
\frac{\partial}{\partial\bar{z}^{\ald}_{N{\setminus}n}|_{p={\ae}_{n}}}
res_{n}z^{\bed}_{N}=0
\eeq
and that
\beq
\label{phi}
res_{n}z^{\ald}_{N}=\epsilon^{\ald \bed}
\frac{\partial}{{\partial}{z}^{\bed}_{N{\setminus}n}|_{p={\ae}_{n}}}
\phi_{N,n}(z_{N{\setminus}n}|_{p={\ae}_{n}})
\eeq
The function $\phi_{N,n}(z_{N{\setminus}1}|_{p={\ae}_{n}})$ is subject 
to a condition that 
the coordinate $z$ has an expansion of the type of (\ref{poly1}),
(\ref{poly2}). This condition requires the function $\phi$ to be
\beq
\phi_{N,n}(z_{N{\setminus}n}|_{p={\ae}_{n}})={\ej}_{n}
e^{(\lambda_{n},u_{N{\setminus}n}|_{{p={\ae}_{n}}})}
\eeq
and hence the residues of the coordinates $z$ Eq.(\ref{phi}) read as
\beq
res_{n}z^{\ald}_{N}=\lambda^{\ald}_{n}{\ej}_{n}
e^{(\lambda_{n},u_{N{\setminus}n}|_{{p={\ae}_{n}}})}
\eeq
and then the coordinates are found to be $z_{N}=z_{\emptyset}+u_{N}$ with 
\begin{eqnarray}
\label{coord}
u^{\ald}_{N}=
\sum_{n{\in}N}\frac{(p,q_{n})^{2}{\lambda}^{\ald}_{n}}
{(p,{\ae}_{n})(q_{n},{\ae}_{n})^{2}}{\ej}_{n}
e^{(\lambda_{n},u_{N{\setminus}n}|_{p={\ae}_{n}})}\nonumber\\
=
\sum_{n{\in}N}\frac{(p,q_{n})^{2}{\lambda}^{\ald}_{n}}
{(p,{\ae}_{n})(q_{n},{\ae}_{n})^{2}}{\ej}_{n}
e^{\sum_{m{\in}N{\setminus}n}
\frac{({\ae}_{n},q_{m})^{2}(\lambda_{n},\lambda_{m})}
{({\ae}_{n},{\ae}_{m})(q_{m},{\ae}_{m})^{2}}{\ej}_{m}
e^{\sum_{l{\in}N{\setminus}m,n}{\ldots}}}
\end{eqnarray}
As soon as one knows the coordinates $z^{\ald}$ one can reconstruct all
the other geometric data (see \cite{RS3}), and I will not reproduce
here all the construction since to know $z^{\ald}$ is sufficient to proceed
in the YM sector.

\section{Gravitational dressing of YM fileds}
In the ptb vierbein $e^{\alpha, \ald}_{N}$  the YM curvature form, $F=dA+A^{2}$,
has four indexes,  $F_{\alpha \ald \beta \bed}$, and, being antisymmetric with respect to 
permutation $(\alpha \ald)\leftrightarrow(\beta \bed)$, decomposes as
\beq
\label{decom}
F_{\alpha \ald \beta \bed}=\varepsilon_{\alpha \beta}
F_{\ald \bed}+\varepsilon_{\ald \bed} F_{\alpha \beta}
\eeq
The first term in the r.h.s. of Eq.~(\ref{decom})
can be identified as an SD component of $F$, the second one - as an anti-SD component
of $F$. Correspondingly, the SD equation can be written as
\beq
\label{SD1}
F_{\alpha  \beta}=0
\eeq
With use of the same constant spinor  $p_{\alpha}$ as in the previous section, one defines
\begin{eqnarray}
\label{1}
A_{\ald}=p^{\alpha}A_{\alpha \ald},\nonumber\\
\dald=p^{\alpha} \partial_{\alpha \ald}, \;
{\rm where} \; \partial_{\alpha \ald}=
{e_{N}}_{\alpha \ald}^{\mu \mud}\frac{\partial}{\partial x^{\mu \mud}},\nonumber\\
\nald=\dald+A_{\ald}
\end{eqnarray}
With these definitions the SD equation (\ref{SD1}) becomes a zero-curvature condition
\beq
\label{SD2}
[\nald, \ned]=0, \; {\rm at \: any} \; p^{\alpha}, \alpha=1,2
\eeq
which is solved as
\beq
\label{gauge}
A_{\ald}=g^{-1}{\dald}g
\eeq
where $g$ is a function of $x^{\alpha \ald}$ and $p^{\alpha}$ with values in the 
complexification of the
gauge group. $g$ must depend on $p^{\alpha}$ in such a way that the resulting
$A_{\ald}$ is a linear function of $p^{\alpha}$, $A_{\ald}=p^{\alpha}A_{\alpha \ald}$ . 
Actually, $g$ is sought for as a homogeneous of 
degree zero rational function of $p^{\alpha}$. Such function necessary has singularities
in the $p^{\alpha}$-space and it is subject to condition of regularity of 
$A_{\ald}$. Then by construction, $A_{\ald}$ is
a homogeneous  of degree one regular rational function of two 
complex  variables $p^{\alpha}, \alpha=1,2$. As such, it is necessary
just linear in $p^{\alpha}$. 

An essential moment is that in the case of {perturbiner} $g^{ptb}$ can only be a
polynomial in the variables ${\ej}_{n}, n{\in}N,{\ej}_{j}, j=1,J$. 
First order in ${\ej}_{j}$ term in $g^{ptb}$
is fixed by the plane wave solution Eq.(\ref{free1})
of the free equation, while the demand of regularity of $A_{\ald}$ fixes  $g^{ptb}$
up to the gauge freedom.

Notice that the only difference with the case of pure YM  is the appearence of
 the ptb vierbein ${e_{N}}_{\alpha \ald}^{\mu \mud}$ in the definition of $\dald$
and hence in the Eq.(\ref{gauge}). Remember that  $g^{ptb}$ is fixed by the
condition of regularity of $A_{\ald}$. Since the ptb vierbein enters Eq.(\ref{gauge}),
it affects the regularity condition and gives rise to the gravitational dressing
of the YM fields. Namely, using the nilpotency of ${\hej}$, one can see that
\beq
\label{induc}
g^{ptb}_{J}=g^{ptb}_{J-1}(1+\chi_{J})
\eeq
where $\chi_{J}$ is of first order in ${\hej}^{J}$ and polynomial in all 
${\hej}^{j}, j=1,J$. (In the Eq.(\ref{induc}) $g^{ptb}_{J}$ 
includes $1, {\ldots}, J$-gluons and  $g^{ptb}_{J-1}$ includes 
$1, {\ldots}, J-1$-gluons.) Then, absolutely parallel to \cite{RS2},
$\chi_{J}$ can be seen to have a simple pole at $p^{\alpha}={\ae}^{\alpha}_{J}$
while its residue at this pole must obey the condition
\beq
\label{res}
[{\dald}({\gpk}(res_{\pk} \chi_{J}){\gpi})]|_{\pk}=0
\eeq  
This last equation in the case of pure YM \cite{RS2} has the only solution
\beq
\label{resym}
({\gpk}(res_{\pk} \chi_{J}){\gpi})=const.{\hej}_{J}
\eeq
while now, in the YM+gravity case it has solution
\beq
\label{resymgra}
({\gpk}(res_{\pk} \chi_{J}){\gpi})=
const.{\hej}_{J}e^{({\lambda}_{J},u_{\pk})}
\eeq
with $u$ from Eq.(\ref{coord}). This defines the gravitational dressing ${\hej}^{gr}_{j}$
of the plane wave ${\hej}_{j}$,
\beq
\label{dressing}
{\hej}^{gr}_{j}={\hej}_{j}e^{({\lambda}_{j},u_{\pk})} 
\eeq
From Eqs.(\ref{induc}),(\ref{resymgra}) it is not difficult to see
that this dressing is the only effect of gravity in $g_{ptb}$, that is
$g^{YM+gravity}_{ptb}$ reads as follows 
\beq
\label{dressing1}
g^{YM+gravity}_{ptb}({\hej}_{j})=g^{YM}_{ptb}({\hej}^{gr}_{j})
\eeq
with $g^{YM}_{ptb}$ as in \cite{RS2}.
For closeness of the presentation, I  give here the resulting expression
for the color ordered highest degree monomial in $g^{YM+gravity}_{ptb}$,
that is for the term in $g^{YM+gravity}_{ptb}$ where all gluonic plane
waves ${\hej}_{j}$ are present and come in a given order, say, in the order
${\hej}_{J}{\hej}_{J-1} {\ldots}{\hej}_{1}$.
\beq
\label{bsol}
g^{ptb}_{J (J, \ldots ,1)}=\frac{(p, q^{J})(\ae^{J}, q^{J-1}) \ldots 
(\ae^{2}, q^{1})}{(p, \ae^{J})(\ae^{J}, \ae^{J-1}) \ldots
(\ae^{2}, \ae^{1})}
\frac{ {\hej}^{gr}_{J}}{( \ae^{J},q^{J})} \ldots \frac{{\hej}^{gr}_{1}}
{( \ae^{1},q^{1})}
\eeq
This is, essentially, a solution of the problem.
Substituting $g^{ptb}$ (\ref{bsol}) into equation (\ref{gauge}) determines 
the perturbiner $A^{ptb}_{\ald}$. The computation is easy and parallel to
the one in the pure YM case \cite{RS2}, so I will not reproduce it here.
Actually, $g_{ptb}$ contains more information than $A_{ptb}$ and is more
useful in applications, such as construction of the gravitationally dressed
Parke-Taylor amplitudes which will be described elsewhere.

{\it Acknowledgments}\\
This work could not be possible without permanent discussions with
A.Rosly, whom I am very much obliged to.
I would also like to thank L.B.Okun for support from RFFI- 96-15-96578
and H.Leutwyler and other members
of ITP at  Bern University for their kind hospitality when this work was done.

\end{document}